\documentclass[twoside,fleqn]{article}
\usepackage{espcrc2}

\usepackage{graphicx}
\usepackage[figuresright]{rotating}

\def\beq{\begin{equation}} 
\def\eeq{\end{equation}}

\def\beqn{\begin{eqnarray}} 
\def\eeqn{\end{eqnarray}}

\def\({\left(} 
\def\){\right)}

\newcommand{\AmS}{{\protect\the\textfont2
  A\kern-.1667em\lower.5ex\hbox{M}\kern-.125emS}}

\newcommand\sss{\mathchoice%
{\displaystyle}%
{\scriptstyle}%
{\scriptscriptstyle}%
{\scriptscriptstyle}%
}

\newcommand\POWHEG{{\tt POWHEG}}
\newcommand\POWHEGBOX{{\tt POWHEG BOX}}

\newcommand\HERWIG{{\tt HERWIG}}

\newcommand\PYTHIA{{\tt PYTHIA}}
\newcommand\ARIADNE{{\tt ARIADNE}}

\newcommand\pt{p_{\sss\rm T}}

\begin{document}

%


\title{The \POWHEGBOX{}}

\author{
        C. Oleari\address{Dipartimento di Fisica "G. Occhialini" and INFN,
	  Sezione di Milano-Bicocca,  
        Universit\`a di Milano-Bicocca, 
        20126 Milano, Italy}
}

\begin{abstract}
We review the key features of \POWHEG{}, a method for interfacing
parton-shower Monte Carlo generators to fixed next-to-leading order QCD
computations. We describe a recently introduced framework, the \POWHEGBOX{},
that allows the automatic \POWHEG{} implementation of any given NLO
calculation. We present a few results for Higgs boson production via vector
boson fusion and $Z+1$~jet production, both processes available in the
\POWHEGBOX{}.
\end{abstract}

\maketitle

\section{The \POWHEG{} method}
In the past two decades, next-to-leading order~(NLO) QCD computations have
become standard tools for phenomenological studies at lepton and hadron
colliders. QCD tests have been mainly performed by comparing NLO results with
experimental measurements, with the latter corrected for detector effects.

On the experimental side, leading order~(LO) calculations, implemented in the
context of general purpose Shower Monte Carlo~(SMC) programs, have been the
main tools used in the analysis. SMC programs include dominant QCD effects at
the leading logarithmic level, but do not enforce NLO accuracy. These
programs were routinely used to simulate background processes and signals in
physics searches. When a precision measurement was needed, to be compared
with a NLO calculation, one could not directly compare the experimental
output with the SMC output, since the SMC does not have the required
accuracy. The SMC output was used, in this case, to correct the measurement
for detector effects, and the corrected result was compared to the NLO
calculation.

In view of the positive experience with QCD tests at the NLO level, it has
become clear that SMC programs should be improved, when possible, with NLO
results. In this way a large amount of the acquired knowledge on QCD
corrections would be made directly available to the experimentalists in a
flexible form that they could easily use for simulations.

The problem of merging NLO calculations with parton shower simulations is
basically that of avoiding overcounting, since the SMC programs do implement
approximate NLO corrections already. 

The features implemented in \POWHEG{} can be summarized as follows:
\begin{itemize}
\item[-] infrared-safe observables have NLO accuracy;
\item[-] collinear emissions are summed at the leading-logarithmic level;
\item[-] the double logarithmic region (i.e.\ soft and collinear gluon
  emission) is treated correctly if the SMC code used for showering has this
  capability.
\end{itemize}
In the case of \HERWIG{}~\cite{Corcella:2000bw,Corcella:2002jc} this last
requirement is satisfied, owing to the fact that its shower is based upon an
angular-ordered branching.

In ref.~\cite{Nason:2004rx} a method, called \POWHEG{} (for Positive Weight
Hardest Emission Generator), was proposed that produces positive-weighted
events, and that is not SMC specific. In the \POWHEG{} method the hardest
radiation is generated first using the exact NLO matrix elements.  The
\POWHEG{} output can then be interfaced to any SMC program that is either
$\pt$-ordered, or allows the implementation of a $\pt$ veto.\footnote{All SMC
  programs compatible with the {\em Les Houches Interface for User
    Processes}~\cite{Boos:2001cv} should comply with this requirement.}
However, when interfacing \POWHEG{} to angular-ordered SMC programs, the
double-log accuracy of the SMC is not sufficient to guarantee the double-log
accuracy of the whole result.  Some extra soft radiation (technically called
vetoed-truncated shower in ref.~\cite{Nason:2004rx}) must also be included in
order to recover double-log accuracy. In fact, angular ordered SMC programs
may generate soft radiation before generating the radiation with the largest
$\pt$, while \POWHEG{} generates it first. When \POWHEG{} is interfaced to
shower programs that use transverse-momentum ordering, the double-log
accuracy should be correctly retained if the SMC is double-log accurate. The
\ARIADNE{} program~\cite{Lonnblad:1992tz} and
\PYTHIA{}~6.4~\cite{Sjostrand:2006za} (when used with the new showering
formalism), both adopt transverse-momentum ordering, in the framework of
dipole-shower algorithm, and aim to have accurate soft resummation
approaches, at least in the limit of large number of colours.

In the \POWHEG{} formalism~\cite{Frixione:2007vw}, the generation of the
hardest emission is performed first, according to the distribution
\beqn
\label{eq:master}
d\sigma&=&\bar{B}\left(\Phi_{B}\right)\,d\Phi_{B}\,\bigg[ \Delta
  \left(p_{\sss\rm T}^{\min}\right) \nonumber \\
&&+   
\frac{R\left(\Phi_{R}\right)}{B\left(\Phi_{B}\right)}\,\Delta 
\left(k_{\sss\rm T}\left(\Phi_{R}\right) \right)\,d\Phi_{\mathrm{rad}}\bigg]\,,
\eeqn
where $B\left(\Phi_{B}\right)$ is the leading order Born contribution and 
\begin{equation}
\label{eq:bbar}
\bar{B}\left(\Phi_{B}\right)=B\left(\Phi_{B}\right)+
\left[V\left(\Phi_{B}\right)+\!\!\int \!\!d\Phi_{\mathrm{rad}}\,
  R\left(\Phi_{R}\right)\right]
\end{equation}
is the NLO differential cross section used to generate the Born variables
($V\left(\Phi_{B}\right)$ and $R\left(\Phi_{R}\right)$ stand respectively for
the virtual and the real corrections), and 
\beqn
&&\mbox{}\hspace{-5mm} \Delta\left(p_{\sss\rm T}\right) = \nonumber \\
&&\exp\left[-\int d\Phi_{\mathrm{rad}}\,
  \frac{R\left(\Phi_{R}\right)}{B\left(\Phi_{B}\right)}
  \,\theta\left(k_{\sss\rm T}\left(\Phi_{R}\right)-p_{\sss\rm T}
  \right)\right]\phantom{aa} 
\eeqn 
is the \POWHEG{} Sudakov form factor.  The transverse momentum of the emitted
particle is here indicated with $k_{\sss\rm T}\left(\Phi_{R}\right)$.  The
cancellation of soft and collinear singularities is understood in the
expression within the square bracket in eq.~(\ref{eq:bbar}).  Partonic events
with hardest emission generated according to eq.~(\ref{eq:master}) are then
showered with a $k_{\sss\rm T}$-veto on following emissions.  For all the
technicalities concerning this and other issues, we refer
to~\cite{Nason:2004rx,Frixione:2007vw}.

Up to now, the \POWHEG{} method, in the context of hadron colliders, has been
applied to $ZZ$ pair hadroproduction~\cite{Nason:2006hf}, heavy-flavour
production~\cite{Frixione:2007nw}, Drell-Yan vector boson
production~\cite{Alioli:2008gx,Hamilton:2008pd}, Higgs boson production via
gluon fusion~\cite{Alioli:2008tz,Hamilton:2009za}, Higgs boson production
associated with a vector boson (Higgs-strahlung)~\cite{Hamilton:2009za},
single-top production~\cite{Alioli:2009je}, Higgs boson production in vector
boson fusion~\cite{Nason:2009ai} and $Z+1$~jet
production~\cite{POWHEG_Zjet}. 

\section{The \POWHEGBOX}

The \POWHEGBOX{}~\cite{Alioli:2010xd} is a computer framework that implements
in practice the theoretical construction of \POWHEG{}. The aim of the
\POWHEGBOX{} is to construct a \POWHEG{} implementation of a NLO process,
given the following ingredients:
\begin{enumerate}
\item The list of all flavour structures of the Born processes.

\item The list of all flavour structures of the real processes.
 
\item  The Born phase space.

\item \label{item:born} The Born squared amplitudes ${\cal B}$, the color
  correlated ones ${\cal B}_{ij}$ and spin correlated ones ${\cal B}_{\mu\nu}$.
 These are common ingredients of NLO calculations performed
  with a subtraction method.
  
\item \label{item:real} The real matrix elements squared for all relevant
  partonic processes.

\item \label{item:virtual} The finite part of the virtual corrections
  computed in dimensional regularization or in dimensional
  reduction.
    
\item The Born colour structures in the limit of a large  number of colours.
\end{enumerate}

With the exception of the virtual corrections, all these ingredients are
nowadays easily obtained. A matrix element program can be used to
obtain~(\ref{item:born}) and~(\ref{item:real}).  The colour-correlated and
spin-correlated Born amplitudes can also be generated automatically.  Recent
progress in the automatization of the virtual cross section calculation may
lead to developments where even the virtual contribution~(\ref{item:virtual})
may be obtained in a painless
way.
Given the ingredients listed above, the \POWHEGBOX{} does all the rest.  It
automatically finds all the singular regions, builds the soft and collinear
counterterms and the soft and collinear remnants, and then generates the
radiation using the \POWHEG{} Sudakov form factor.

\section{Higgs boson production in vector-boson fusion}
Higgs boson production via vector-boson fusion~(VBF) is expected to provide a
copious source of Higgs bosons in $pp$-collisions at the Large Hadron
Collider~(LHC) at CERN. It can be visualized as the inelastic scattering of
two quarks (antiquarks), mediated by $t$-channel $W$ or $Z$ exchange, with
the Higgs boson radiated off the weak bosons.  It represents (after gluon
fusion) the second most important production process for Higgs boson
studies~\cite{CMS,ATLAS}.  Once the Higgs boson has been found and its mass
determined, the measurement of its couplings to gauge bosons and fermions
will be of main interest~\cite{Zeppenfeld:2000td,Duhrssen:2004cv}.  Here VBF
will play a central role since it will be observed in the
$H\to\tau\tau$~\cite{Rainwater:1998kj,Plehn:1999xi}, $H\to
WW$~\cite{Rainwater:1999sd,Kauer:2000hi} and
$H\to\gamma\gamma$~\cite{Rainwater:1997dg} channels. This multitude of
channels allows to probe the different Higgs boson couplings.  
In addition, in order to distinguish the VBF Higgs boson signal from
backgrounds, stringent cuts are required on the Higgs boson decay products as
well as on the two forward quark jets which are characteristic for VBF. The
efficiency of these cuts has to be evaluated on the basis of the most updated
simulation tools and experimental inputs.

\begin{figure}[t!] 
  \begin{center}
\includegraphics[width=1\columnwidth]{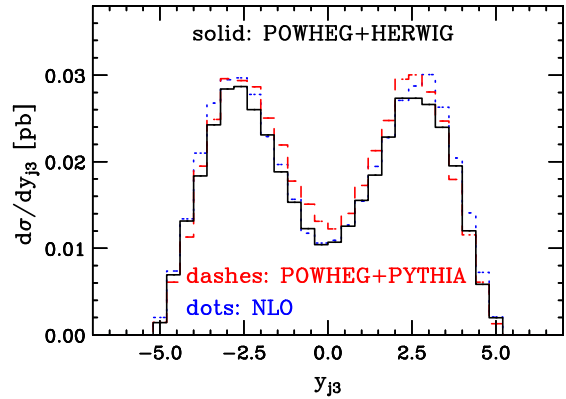}
\caption{
\label{fig:y3}
Rapidity $y_{j_3}$ of the third hardest jet, i.e.~the
one with highest $\pt$ after the two tagging jets.}
  \end{center}
\end{figure}

An additional feature characterizing VBF Higgs boson production is the fact
that, at leading order, no colored particle is exchanged in the $t$ channel
so that no $t$-channel gluon exchange is possible at NLO, once we neglect the
small contribution due to equal-flavour quark scattering with
$t\leftrightarrow u$ channel interference.  The different gluon radiation
pattern expected for Higgs boson production via VBF compared to its major
backgrounds ($t\bar{t}$ production, QCD $WW + 2$~jet and QCD $Z+2$~jet
production) is at the core of the central-jet veto proposal, both for
light~\cite{Kauer:2000hi} and heavy~\cite{Barger:1994zq} Higgs boson
searches.  A veto of any additional jet activity in the central-rapidity
region is expected to suppress the backgrounds more than the signal, because
the QCD backgrounds are characterized by quark or gluon exchange in the
$t$-channel. The exchanged partons, being colored, are expected to radiate
off more gluons.

For the analysis of the Higgs boson coupling to gauge bosons, Higgs
boson~+~2~jet production via gluon fusion may also be treated as a background
to VBF.  When the two jets are separated by a large rapidity interval, the
scattering process is dominated by gluon exchange in the $t$-channel.
Therefore, like for the QCD backgrounds, the bremsstrahlung radiation is
expected to occur everywhere in rapidity. An analogous difference in the
gluon radiation pattern is expected in $Z+2$~jet production via VBF fusion
versus QCD production~\cite{Rainwater:1996ud}.  In order to analyze this
feature, we have plotted in fig.~\ref{fig:y3} the rapidity of the third jet,
the one with highest $\pt$ after the two tagging jets.  The conclusion that
can be drawn from the figure is that the third jet in VBF prefers to be
emitted close to one of the tagging jets, while, in gluon fusion, it is
emitted anywhere in the rapidity region between the tagging jets. Thus, at
least with regard to the hard radiation of a third jet, the analysis of
refs.~\cite{DelDuca:2004wt,DelDuca:2006hk,Andersen:2008gc} is confirmed.
The distributions, obtained using \POWHEG{} interfaced to \HERWIG{} and
\PYTHIA{}, are very similar and turn out to be well modeled by the respective
distributions of the NLO jet.

\begin{figure}[t!] 
  \begin{center}
\includegraphics[width=1\columnwidth]{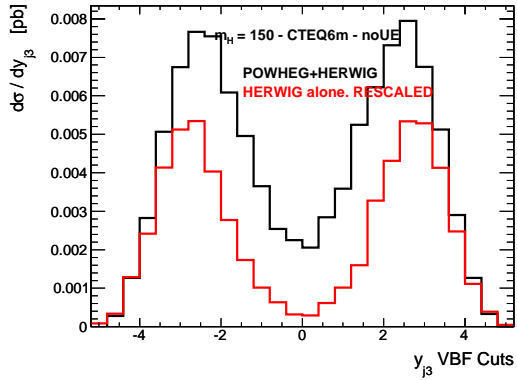}
\caption{
\label{fig:yj_3_WBFcuts_HW}
Comparison between the rapidity distribution of the third jet, $y_{j3}$,
obtained using \POWHEG{} followed by the shower of \HERWIG{} (in black), and
\HERWIG{} alone (in dark grey). The two curves have been normalised such that
the total cross sections agree at NLO level.}
  \end{center}
\end{figure}

\begin{figure}[t!] 
  \begin{center}
\includegraphics[width=1\columnwidth]{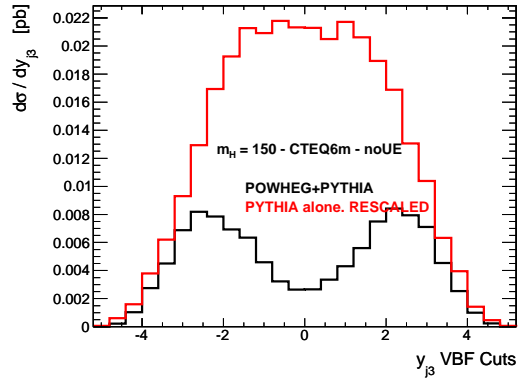}
\caption{
\label{fig:yj_3_WBFcuts_PY}
Comparison between the rapidity distribution of the third jet, $y_{j3}$,
obtained using \POWHEG{} followed by the shower of \PYTHIA{} (in black), and
\PYTHIA{} alone (in dark grey). The two curves have been normalized such that
the total cross sections agree at NLO level.}
  \end{center}
\end{figure}

For comparison, in figs.~\ref{fig:yj_3_WBFcuts_HW}
and~\ref{fig:yj_3_WBFcuts_PY} we have plotted the same distribution as
obtained with \HERWIG{} and \PYTHIA{} alone. Both the two plots are
normalized such that the total cross section is correct at NLO.  In
fig.~\ref{fig:yj_3_WBFcuts_HW}, it is clear thatthe third jet produced by
\HERWIG{} underestimates the actual third-jet rapidity, even if the shape is
similar. \PYTHIA{} has instead a completely different behavior: here not only
the normalization is wrong, but the predicted $y_{j3}$ distribution is not
correct.  This is of no surprise, since both \HERWIG{} and \PYTHIA{} produce
(and resum) correctly only jets in the collinear region (collinear with
respect to the two tagging jets), so that we do not expect any
distribution involving the third jet to be correct.

\section{${\bf Z+1}$~jet production}
In the early LHC phase, $Z+1$~jet production is a promising process for jet
calibration. Both $Z$ and $W$ production in association with jets are
important sources of missing energy signals, and $W$ in association with jets
is an important background to many new physics searches. The $Z$ production
process, with the $Z$ decaying into leptons, gives a distinct signature that
allows to perform stringent tests of the production mechanism. These tests
have been performed extensively at the Tevatron, both at
CDF~\cite{Aaltonen:2007cp}  and at
D0~\cite{Abazov:2006gs,Abazov:2008ez,Abazov:2009av,Abazov:2009pp}.
They are carried out by correcting the measured quantities to the particle
level, according to the recommendations developed in the 2007 Les Houches
workshop~\cite{Buttar:2008jx}, and then compared to NLO theoretical
calculations, corrected for showering and underlying event effects. These
corrections, in turn, are extracted from SMC programs.

\begin{figure}[t!] 
  \begin{center}
\includegraphics[width=1\columnwidth]{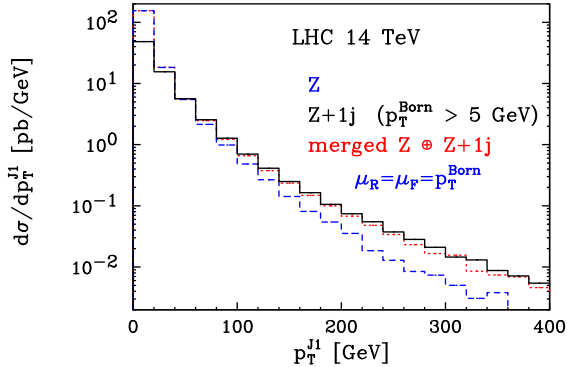}
\caption{
\label{fig:ZjZ_ptj1}
The $\pt$ distribution of the first jet, in single $Z$ production (blue
dashed curve), in $Z+1$~jet production (black solid curve) and of the merged
sample (red dotted curve).}
  \end{center}
\end{figure}

Very recently we have implemented $Z+1$~jet in the \POWHEGBOX{}. As a matter
of fact, the \POWHEGBOX{} framework was developed with the $Z+1$~jet process
as its first example process. It is clear that, by using this tool, the
comparison of the theoretical prediction with the experimental results is
eased considerably, and is also made more precise. Rather than estimating
shower and underlying event corrections, using a parton shower program, these
corrections can now be applied directly to the hard process in question,
yielding an output that can be compared to the experimental results at the
particle level. We have carry out this task and compared our results to the
Tevatron findings.  We remark, however, that a further improvement to this
study could be carried out, by using the \POWHEG{} program to generate the
events that are fed into the detector simulation, and are directly compared
to raw data. We are not, of course, in a position to perform such a task,
that should instead be carried out by the experimental collaborations.

\begin{figure}[t!] 
  \begin{center}
\includegraphics[width=1\columnwidth]{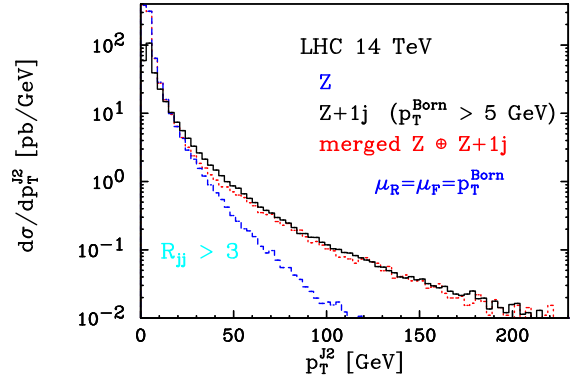}
\caption{
\label{fig:ZjZ_ptj2_R3}
The $\pt$ distribution of the second jet, in single $Z$ production (blue
dashed curve), in $Z+1$~jet production (black solid curve) and of the merged
sample (red dotted curve), when an angular separation $R_{jj}>3$ is imposed
between the two leading jets.}
  \end{center}
\end{figure}

Since a \POWHEG{} implementation of $Z$ production has already appeared in
the literature~\cite{Alioli:2008gx}, a natural question arises, weather we
can now, with the new $Z+1$~jet calculation, build a merged sample of the two
programs, capable of describing $Z$ production in the low, as well as in the
high transverse-momentum region. We have presented, for the first time, a
minimalistic approach to such a merging, by putting together independent
samples of $Z$ and $Z+1$~jet events in the most straightforward way,
described in details in a forthcoming paper~\cite{POWHEG_Zjet}.

In fig.~\ref{fig:ZjZ_ptj1} we have plotted the $\pt$ distribution of the
first jet, in single $Z$ production (blue dashed curve), in $Z+1$~jet
production (black solid curve) and of the merged sample (red dotted
curve). As can be seen in the figure, the merged distribution smoothly
interpolates between the single $Z$ behavior at small $\pt$, where the
Sudakov form factor plays an important role in resumming the divergent
collinear singularities, and the $Z+1$~jet curve, at high $\pt$, where the
presence of the second hard jet in the $Z+1$~jet sample, hardens the
distribution of the first jet.  Similar interpolating behavior is seen for
the $\pt$ of the second jet, in fig.~\ref{fig:ZjZ_ptj2_R3}. Here, an angular
separation $R_{jj}>3$ is imposed between the two leading jets, in order to
separate them further to better illustrate the behavior of the merged curve.

\section{Acknowledgments}
The results presented in this talk have been obtained in collaboration
with S.~Alioli, P.~Nason and E.~Re.

\bibliography{paper}

\begin{thebibliography}{10}

\bibitem{Corcella:2000bw}
G. Corcella et~al.,
\newblock JHEP 01 (2001) 010, hep-ph/0011363.

\bibitem{Corcella:2002jc}
G. Corcella et~al.,
\newblock (2002), hep-ph/0210213.

\bibitem{Nason:2004rx}
P. Nason,
\newblock JHEP 11 (2004) 040, hep-ph/0409146.

\bibitem{Boos:2001cv}
E. Boos et~al.,
\newblock (2001), hep-ph/0109068.

\bibitem{Lonnblad:1992tz}
L. Lonnblad,
\newblock Comput. Phys. Commun. 71 (1992) 15.

\bibitem{Sjostrand:2006za}
T. Sjostrand, S. Mrenna and P. Skands,
\newblock JHEP 05 (2006) 026, hep-ph/0603175.

\bibitem{Frixione:2007vw}
S. Frixione, P. Nason and C. Oleari,
\newblock JHEP 11 (2007) 070, 0709.2092.

\bibitem{Nason:2006hf}
P. Nason and G. Ridolfi,
\newblock JHEP 08 (2006) 077, hep-ph/0606275.

\bibitem{Frixione:2007nw}
S. Frixione, P. Nason and G. Ridolfi,
\newblock JHEP 09 (2007) 126, 0707.3088.

\bibitem{Alioli:2008gx}
S. Alioli et~al.,
\newblock JHEP 07 (2008) 060, 0805.4802.

\bibitem{Hamilton:2008pd}
K. Hamilton, P. Richardson and J. Tully,
\newblock JHEP 10 (2008) 015, 0806.0290.

\bibitem{Alioli:2008tz}
S. Alioli et~al.,
\newblock JHEP 04 (2009) 002, 0812.0578.

\bibitem{Hamilton:2009za}
K. Hamilton, P. Richardson and J. Tully,
\newblock JHEP 04 (2009) 116, 0903.4345.

\bibitem{Alioli:2009je}
S. Alioli et~al.,
\newblock JHEP 09 (2009) 111, 0907.4076.

\bibitem{Nason:2009ai}
P. Nason and C. Oleari,
\newblock JHEP 02 (2010) 037, 0911.5299.

\bibitem{POWHEG_Zjet}
S. Alioli et~al.,
\newblock {\rm to appear soon.}

\bibitem{Alioli:2010xd}
S. Alioli et~al.,
\newblock JHEP 06 (2010) 043, 1002.2581.

\bibitem{CMS}
{CMS Collaboration},
\newblock Report CMS/LHCC/2006-021, CMS TDR 8.2.

\bibitem{ATLAS}
{ATLAS Collaboration},
\newblock Report CERN/LHCC/99-15 (1999).

\bibitem{Zeppenfeld:2000td}
D. Zeppenfeld et~al.,
\newblock Phys. Rev. D62 (2000) 013009, hep-ph/0002036.

\bibitem{Duhrssen:2004cv}
M. Duhrssen et~al.,
\newblock Phys. Rev. D70 (2004) 113009, hep-ph/0406323.

\bibitem{Rainwater:1998kj}
D.L. Rainwater, D. Zeppenfeld and K. Hagiwara,
\newblock Phys. Rev. D59 (1999) 014037, hep-ph/9808468.

\bibitem{Plehn:1999xi}
T. Plehn, D.L. Rainwater and D. Zeppenfeld,
\newblock Phys. Rev. D61 (2000) 093005, hep-ph/9911385.

\bibitem{Rainwater:1999sd}
D.L. Rainwater and D. Zeppenfeld,
\newblock Phys. Rev. D60 (1999) 113004, hep-ph/9906218.

\bibitem{Kauer:2000hi}
N. Kauer et~al.,
\newblock Phys. Lett. B503 (2001) 113, hep-ph/0012351.

\bibitem{Rainwater:1997dg}
D.L. Rainwater and D. Zeppenfeld,
\newblock JHEP 12 (1997) 005, hep-ph/9712271.

\bibitem{Barger:1994zq}
V.D. Barger, R.J.N. Phillips and D. Zeppenfeld,
\newblock Phys. Lett. B346 (1995) 106, hep-ph/9412276.

\bibitem{Rainwater:1996ud}
D.L. Rainwater, R. Szalapski and D. Zeppenfeld,
\newblock Phys. Rev. D54 (1996) 6680, hep-ph/9605444.

\bibitem{DelDuca:2004wt}
V. Del~Duca, A. Frizzo and F. Maltoni,
\newblock JHEP 05 (2004) 064, hep-ph/0404013.

\bibitem{DelDuca:2006hk}
V. Del~Duca et~al.,
\newblock JHEP 10 (2006) 016, hep-ph/0608158.

\bibitem{Andersen:2008gc}
J.R. Andersen, V. Del~Duca and C.D. White,
\newblock JHEP 02 (2009) 015, 0808.3696.

\bibitem{Aaltonen:2007cp}
CDF - Run II, T. Aaltonen et~al.,
\newblock Phys. Rev. Lett. 100 (2008) 102001, 0711.3717.

\bibitem{Abazov:2006gs}
D0, V.M. Abazov et~al.,
\newblock Phys. Lett. B658 (2008) 112, hep-ex/0608052.

\bibitem{Abazov:2008ez}
D0, V.M. Abazov et~al.,
\newblock Phys. Lett. B669 (2008) 278, 0808.1296.

\bibitem{Abazov:2009av}
D0, V.M. Abazov et~al.,
\newblock Phys. Lett. B678 (2009) 45, 0903.1748.

\bibitem{Abazov:2009pp}
D0, V.M. Abazov et~al.,
\newblock Phys. Lett. B682 (2010) 370, 0907.4286.

\bibitem{Buttar:2008jx}
C. Buttar et~al.,
\newblock (2008), 0803.0678.

\end{thebibliography}

\end{document}